\documentclass{80SA}
\usepackage{times}

\title{Magnetoresistance and Phase Diagram of Thin-Film UNi$_2$Al$_3$}

\author{Marc Scheffler$^{1}$\thanks{E-mail address: scheffl@pi1.physik.uni-stuttgart.de}, Eva Rose$^{1}$, Julia P. Ostertag$^{1}$, Karl Schrem$^{1}$, Katrin Steinberg$^{1}$, Martin Dressel$^{1}$, Martin Jourdan$^{2}$ 
}
\inst{$^{1}$1. Physikalisches Institut, Universit\"at Stuttgart, D-70550 Stuttgart, Germany\\
$^{2}$Institut f\"ur Physik, Johannes Gutenberg Universit\"at, D-55099 Mainz, Germany 
}

\abst{We study the dc resistivity of UNi$_2$Al$_3$ thin films as a function of temperature and magnetic field. We focus on the temperature range around the antiferromagnetic transition ($T_N \approx$ 4~K in zero applied field). From a clear signature of $T_N$ in the dc resistance along the crystallographic $a$-direction, we extract the shape of the magnetic phase diagram. Here we find quantitative differences in comparison to previous studies on bulk crystals.}

\kword{UNi$_2$Al$_3$, UPd$_2$Al$_3$, phase diagram, magnetoresistance}

\begin{document}
\maketitle


Amongst the uranium-based heavy-fermion compounds, UNi$_2$Al$_3$ and UPd$_2$Al$_3$ play an important role:\cite{Geibel1991a,Geibel1991b,Pfleiderer2009} both exhibit heavy-fermion superconductivity in an antiferromagnetic phase. From this observation, when considered together with the identical, hexagonal crystal structure, one might expect very similar behavior for both compounds, but this is not the case. The critical temperatures $T_N$ for antiferromagnetism and $T_c$ for superconductivity differ by a factor of 3 and 2, respectively ($T_N$~$\approx$ 4~K and $T_c$~= 1~K for UNi$_2$Al$_3$; $T_N$~= 14~K and $T_c$~= 2~K for UPd$_2$Al$_3$), and also the effective magnetic moment per uranium atom in the antiferromagnetic phase differs considerably: 0.85$\mu_B$ for UPd$_2$Al$_3$,\cite{Krimmel1992} and approximately 0.2$\mu_B$ for UNi$_2$Al$_3$.\cite{Schroeder1994,Lussier1997} Most remarkable is the difference of the antiferromagnetic order: UPd$_2$Al$_3$ has a simple, commensurate ordering wave vector,\cite{Krimmel1992} whereas the wave vector of UNi$_2$Al$_3$ is incommensurate.\cite{Schroeder1994,Lussier1997} According to detailed neutron studies in zero magnetic field, the antiferromagnetic order of UNi$_2$Al$_3$ is an amplitude modulated magnetization wave with the magnetic moments in the basal plane.\cite{Hiess2001}
The magnetic phase diagram of UNi$_2$Al$_3$ was studied previously by means of dc susceptibility, dc resistivity, magnetization, and specific heat.\cite{Sullow1997,Tateiwa1998}

Given the important role of UPd$_2$Al$_3$ for the understanding of unconventional superconductivity in heavy-fermions,\cite{Jourdan1999,Sato2001} it is unfortunate that the experimental situation of UNi$_2$Al$_3$ is still much less established than that of UPd$_2$Al$_3$. One reason for this is that the growth of single-crystalline samples of UNi$_2$Al$_3$ is substantially more difficult than of UPd$_2$Al$_3$.
This restriction does not hold for those experiments, where thin-film samples can be employed or are even advantageous, e.g.\ different types of spectroscopy such as tunneling,\cite{Jourdan1999} optical,\cite{Ostertag2010} or microwave,\cite{Scheffler2005c,Scheffler2006} as well as anisotropic dc resistivity measurements.\cite{Jourdan2004a} Another promising recent prospect of heavy-fermion thin-film growth is the design and control of epitaxial multilayers.\cite{Shishido2010}

We have grown thin films of UNi$_2$Al$_3$ using a molecular beam epitaxy setup: the three constituent elements were evaporated separately (with individually controlled rates) and deposited onto heated YAlO$_3$(112) substrates.\cite{Jourdan2004a,Zakharov2005} The high quality of these thin films is evident from x-ray diffraction as well as the dc resistivity, which corresponds to those of single crystals. The magnetic properties of thin films are much harder to study than those of bulk samples, but using resonant magnetic X-ray scattering it was shown that thin films of the type presented here exhibit the same magnetic order as single crystals.\cite{Jourdan2005}

The films grow in such a way that both the crystallographic $a$- and $c$-axes lie in the film plane and are easily accessible for transport experiments.\cite{Jourdan2004a,Foerster2007,Scheffler2007} 
In the present study, we compare two UNi$_2$Al$_3$ thin films of different thickness: sample \#1 is 150~nm thick and was studied already previously.\cite{Foerster2007} Sample \#2 is only 62~nm thin, which is particularly advantageous for transmission experiments in optics.\cite{Ostertag2010} In the present experiment we concentrate on the dc resistivity along the $a$-axis as a function of temperature and external magnetic field. Here we have performed conventional 4-probe measurements. 

\begin{figure}
\begin{center}
\includegraphics{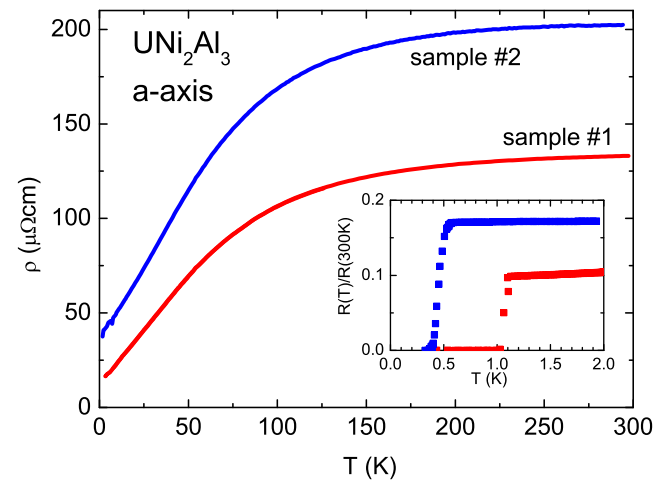}
\end{center}
\caption{Temperature-dependent resistivity along the $a$-axis for UNi$_2$Al$_3$ thin films. Thickness of samples \#1 and \#2 are 150~nm and 62~nm, respectively. The inset shows the resistance around the superconducting transition.}
\label{FigRhoVsT}
\end{figure}

The temperature-dependent resistivity in zero external field is shown for both samples in Fig.\ \ref{FigRhoVsT}: the overall temperature dependence is typical for heavy-fermion materials and known from previous studies on UNi$_2$Al$_3$:\cite{Geibel1991a,Jourdan2004a,Sato1996} upon cooling from room temperature, the resistivity decreases only slightly until around 100~K, where the resistivity smoothly bends over to a steeper descent. Finally, at temperatures around or below 1~K, the resistivity vanishes in the superconducting state. The superconducting transition of our samples is depicted in the inset of Fig.\ \ref{FigRhoVsT}: defining $T_c$ as the midpoint of the resistive transition, we find a $T_c$ of 1.07~K for sample \#1 and 0.46~K for sample \#2. The residual resistivity ratio (RRR) amounts to 9.2 for sample \#1 and 5.9 for sample \#2, respectively. The thicker film \#2 thus is comparable to bulk UNi$_2$Al$_3$ samples.\cite{Geibel1991a,Sato1996} Although $T_c$ and RRR for sample \#2 are reduced compared to sample \#1 and to bulk samples, we want to stress that also this sample is of extremely high quality, if one considers the thickness of only 62~nm. For much thicker films of UPd$_2$Al$_3$, a mean free path of approximately 60~nm was determined,\cite{Scheffler2005c} and similar values will hold for UNi$_2$Al$_3$.\cite{Scheffler2006} Since the film thickness sets a natural upper limit for the mean free path, our very thin film \#2 must have a shorter mean free path, and as a result RRR and $T_c$ are reduced.

\begin{figure}
\begin{center}
\includegraphics{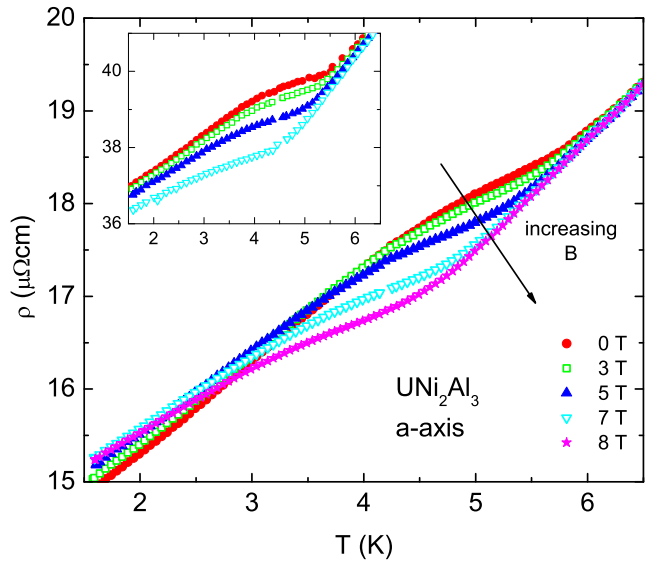}
\end{center}
\caption{(Color online) Temperature-dependent resistivity (along $a$-axis) of thin UNi$_2$Al$_3$ films at temperatures around the antiferromagnetic transition, measured for different external magnetic field strengths. The magnetic field is applied along the $a$-axis, in the film plane. The main plot shows data of sample \#1 (150~nm thick), the inset data of sample \#2 (62~nm thick).}
\label{FigRhoVsTandB}
\end{figure}

\begin{figure}
\begin{center}
\includegraphics{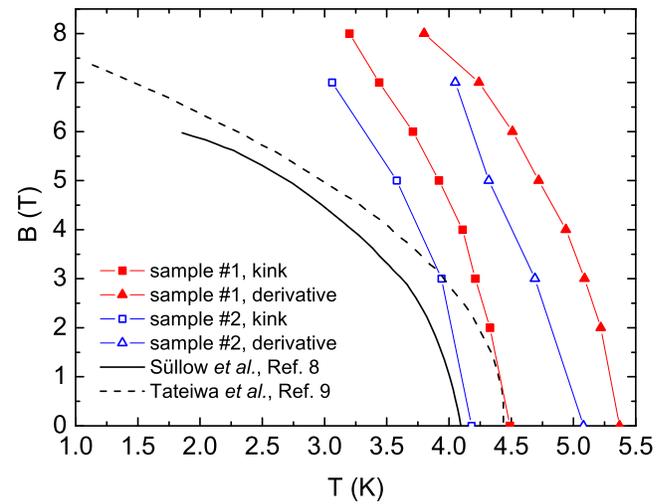}
\end{center}
\caption{(Color online) Phase diagram deduced from the temperature-dependent measurements in Fig.\ \ref{FigRhoVsTandB}. The squares and the triangles correspond to the lower kink in the resistivity and the minimum in the derivative of the resistivity, respectively. The full and dashed black lines represent this phase boundary as determined by S\"ullow \textit{et al.}\cite{Sullow1997} and Tateiwa \textit{et al.},\cite{Tateiwa1998} respectively.}
\label{FigPhaseDiagram}
\end{figure}

In Fig.\ \ref{FigRhoVsTandB}, we present the dc resistivity for temperatures around the antiferromagnetic transition. As observed previously, a plateau-like structure in the resistivity curve is the signature of $T_N$ in the $a$-axis resistivity.\cite{Sullow1997,Jourdan2004a,Sato1996} 
With increasing applied magnetic field, this temperature range with reduced slope moves toward lower temperatures. S\"ullow \textit{et al.} took advantage of this effect to determine the phase boundary of the antiferromagnetic state in temperature-field-plane,\cite{Sullow1997} and we will follow this procedure here as well. The resulting phase diagram is plotted for both samples in Fig.\ \ref{FigPhaseDiagram}.

One unclear aspect in this context is the following: which point of the anomaly of the temperature-dependent resistivity heralds the N\'eel temperature? S\"ullow \textit{et al.} used the minimum of the derivative of resistivity versus temperature, $d \rho(T)/d T$, i.e.\ the minimal slope in $\rho(T)$. But from comparison of data obtained from resonant magnetic X-ray scattering\cite{Jourdan2005} and dc measurements on thin films, we conclude that rather the lower kink of the plateau indicates $T_N$. In Fig.\ \ref{FigPhaseDiagram}, we show both points for our measurements. For comparison, we also plot the phase boundary determined by S\"ullow \textit{et al.}\cite{Sullow1997} from resistivity and susceptibility measurements and by Tateiwa \textit{et al.}\cite{Tateiwa1998} from specific heat and magnetization measurements. In contrast to our thin-film experiments, those data were obtained on single crystals. Surprisingly, we find that the N\'eel temperature of our thin films decreases much less with increasing applied field than for those previous studies. 
If we compare the two films of different thickness, the transition temperature is smaller for the thinner sample. This suggests that with increasing thickness the antiferromagnetic state becomes more robust. But this is at odds with the even lower $T_N$ which was previously found for single crystals.

\begin{figure}
\begin{center}
\includegraphics{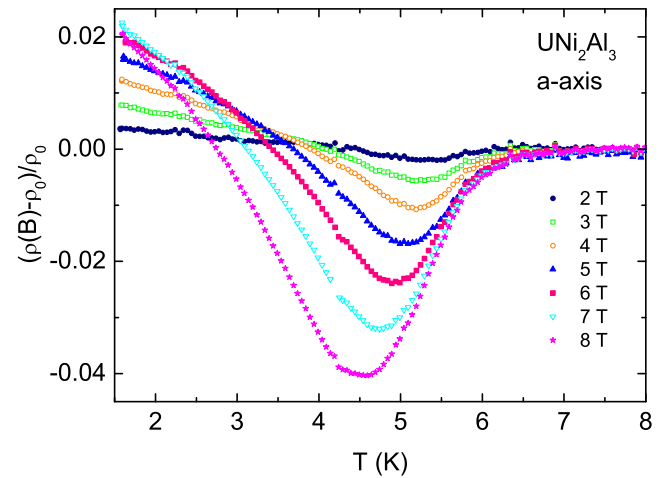}
\end{center}
\caption{(Color online) Temperature-dependent magnetoresistance along the $a$-axis for UNi$_2$Al$_3$ thin film (150~nm thick) for different magnetic fields applied along the $a$-axis.}
\label{FigMR}
\end{figure}

Another difference between our thin film results and the previous single crystal measurements\cite{Sullow1997} becomes more obvious in Fig.\ \ref{FigMR}, where we plot the magnetoresistance for different magnetic fields as a function of temperature. As expected and consistent with the single-crystal result, the magnetoresistance is particularly pronounced for the temperatures around $T_N$, where the suppression of $T_N$ with increasing field leads to substantial negative magnetoresistance. But furthermore we find a positive magnetoresistance at high fields and temperature much lower than $T_N$. This is already evident from the crossing of the curves in the main plot of Fig.\ \ref{FigRhoVsTandB} at temperatures below 4~K.
On the other hand, at temperatures above the plateau in the dc resistivity (\textit{i.e.} above $T_N$) we do not resolve any magnetoresistance; thus any possibly present magnetoresistance at these temperatures is much smaller than 1\%. This is in stark contrast to the single-crystal measurements:\cite{Sullow1997} also well above the plateau, the magnetoresistance there exceeds 6\% at 6~T. This difference could possibly be explained with the dominant scattering mechanisms in our sample: as mentioned above, surface scattering plays an important role, and also other defects probably contribute substantially.\cite{Scheffler2005c} These scattering centers do not depend on temperature or magnetic field, and therefore they have strong constant contributions to the overall resistance which are not affected by magnetic fields. Thus, they cause a large background for any other, magnetic-field-dependent scattering mechanisms which cause the magnetoresistance, such as spin-disorder scattering.\cite{Foerster2007}

\begin{figure}
\begin{center}
\includegraphics{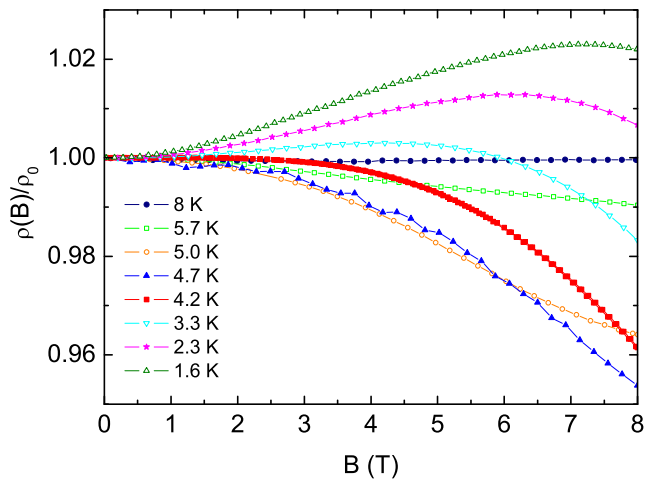}
\end{center}
\caption{(Color online) Normalized resistivity $\rho$($B$)/$\rho_0$=$\rho$($B$)/$\rho$($B$=0) of the 150~nm thick UNi$_2$Al$_3$ film for different temperatures. Current and magnetic field are applied along the crystallographic $a$-axis, in the film plane.}
\label{FigFieldSweeps}
\end{figure}

Finally, we have performed field sweeps at fixed temperature. The resulting normalized resistivity $\rho$($B$)/$\rho_0$=$\rho$($B$)/$\rho$($B$=0) is shown in Fig.\ \ref{FigFieldSweeps} for the thicker sample \#1. These data is consistent with the above: for temperatures clearly above $T_N$ (here: 8~K), we do not resolve any magnetoresistance. At temperatures around 5~K, we find a substantial negative magnetoresistance due to the suppression of $T_N$ with increasing field; and at much lower temperatures the magnetoresistance is positive. These results are very similar to a previous study,\cite{Foerster2007} where the magnetoresistance was studied for current along $a$-axis, but with the magnetic field applied perpendicular to the thin film.

Considering both the absolute value of the magnetoresistance and the field-dependence of the N\'eel temperature determined from our experiments, we obtain a still incomplete picture when compared to the single-crystal results:\cite{Sullow1997} the rather small absolute value of the magnetoresistance for our measurements can be explained by the thin thickness of our samples. Here surface scattering will be a crucial contribution to the overall resistance and does not depend on the applied magnetic field, thus reducing any magnetoresistive effects. The much higher transition 
temperature of the antiferromagnetic phase at large external fields (compared to the single-crystal data) on the other hand is still not understood. 

\section*{Acknowledgements}
We thank S.\ S\"ullow for helpful discussions. This work was supported by the Deutsche Forschungsgemeinschaft (DFG).



\end{document}